\documentclass[11pt]{article}
\usepackage{graphicx}
\usepackage{moriond,epsfig}
\usepackage{fancybox}

\def\Journal#1#2#3#4{{#1} {\bf #2}, #3 (#4)}

\def\PLB{{\em Phys. Lett.}  B}
\def\PRL{\em Phys. Rev. Lett.}
\def\PRD{{\em Phys. Rev.} D}

\def\be{\begin{equation}}
\def\ee{\end{equation}}
\def\bea{\begin{eqnarray}}
\def\eea{\end{eqnarray}}

\begin{document}
\vspace*{4cm}
\title{GEONEUTRINO ANALYSIS IN KamLAND: INPUT AND DESIDERATA}

\author{ G.L. FOGLI, E. LISI, A. PALAZZO, and A.M. ROTUNNO $^\ddagger$ ~
\footnote[0]{$\ddagger$ ({Speaker. E-mail: \tt annamaria.rotunno@ba.infn.it})}}

\address{Dipartimento di Fisica and Sezione INFN, Via Amendola 173,\\
Bari I-70126, Italy}

\maketitle\abstracts{
The Kamioka Liquid scintillator Anti-Neutrino Detector (KamLAND) 
 is collecting antineutrino events generated by
nuclear reactors and by Thorium and Uranium decay in the Earth interior. 
We comment on a systematic approach to the evaluation of the geo-neutrino
contribution and of its uncertainties in KamLAND, taking into account
geophysical and geochemical indications, 
estimates, and data. The results can help to improve both the neutrino
oscillation analysis and the knowledge of the Earth interior.
Input and desiderata for future geoneutrino analyses are identified.
}

\newpage

\section{Introduction}

The Earth surface radiates about 40 TW of heat. About 40\% of this power
energy ($\sim $ 16 TW) is believed to have radiogenic origin, mainly from 
$^{238}$U, $^{232}$Th, and $^{40}$K decays inside the crust and mantle of the Earth
(see, e.g.  \cite{Fiorentini}).
The radiogenic heat is therefore an essential component of the present dynamics of our planet.
These phenomena could be directly studied by detecting the antineutrinos coming from 
 $\beta $-decays of U, Th, K, often called terrestrial antineutrinos, or "geoneutrinos"
($\bar{\nu}_{geo}$).  

The recent results from the Kamioka Liquid scintillator Anti-Neutrino Detector (KamLAND)\cite{KL}
experiment have led to a significant progress in neutrino physics. The observed disappearance of 
reactor antineutrinos is in agreement with the so-called LMA solution of the solar neutrino
problem. Alternative oscillation solutions  are ruled out with high
confidence \cite{Fogli1}.
Geo-neutrino events from  $^{232}$Th and $^{238}$U decays are accessible
for the first time  in KamLAND (those from $^{40}$K decays are below the 
experimental threshold for detection), thus opening a new field in geophysics.

Since $^{232}$Th and $^{238}$U  antineutrino fluxes in KamLAND are weighted by 
the inverse squared distance 1/L$^2$, and 
since $^{232}$Th and $^{238}$U  and $^{40}$K are more abundant in the crust than in the mantle,
some input on the relative $^{232}$Th, $^{238}$U  (and $^{40}$K) abundances in different Earth
reservoirs is needed to make sense of future geoneutrino data. The present work illustrates the importance of 
geochemical studies and inputs, as   necessary and useful tools to shed light on what
we really know about such abundances and on what we expect to know from $^{232}$Th and $^{238}$U  
$\bar{\nu}_{geo}$  data. Furthermore, we analize the impact of KamLAND on the geoneutrino physics.
The results can help to improve both the neutrino oscillation analysis, and the knowledge of the
Earth interior.

\section{Geochemical input for data analysis}

The usually advertised goal of $\bar{\nu}_{geo}$  detection is to measure the Earth radiogenic heat. However, even if the $^{232}$Th and $^{238}$U  
components were known with no error, there would be intrinsic limitations to this goal. We discuss
the most important, as follows. First of all, the 
 $^{40}$K component is  unmeasurable (in KamLAND) and must be inferred from 
			K/U or K/Th ratios estimates. Actually, the 
  K/U and K/Th bulk ratios are not constrained by meteoritic data, since $^{40}$K  is geochemically
			"volatile", namely its condensation temperature is lower than $^{232}$Th and $^{238}$U.
Crust and mantle sampling data, combined with geochemical arguments, are unlikely to reduce
			the K/U and K/Th ratios uncertainty below, say, 10 $\div $ 15 \% \cite{Jochum}, and 
 there might also be a significant amount of potassium (but not $^{232}$Th and $^{238}$U) in the Earth's
			core \cite{rama}. Finally,
 the usually quoted Earth's heat flux, 44$\pm $1 TW \cite{Pollack} might be severely overestimeted by oceanic 
           component systematics and could be as low as 31$\pm $1 TW \cite{Hofmeister}.
These uncertainties set intrinsic limitations to the determination of the radiogenic fraction
of the Earth's heat flux.

The measure of the  $^{232}$Th and $^{238}$U geoneutrino fluxes  appears nevertheless useful  in geophysics,
for the following reasons. Bulk Th/U ratio in the Earth should be close to meteoritic values: 
(Th/U)$_{Earth}$ = (Th/U)$_{chondritic}\sim$ 3.8. There are no geochemical nor cosmochemical arguments 
against this guess. However, Th is more easily partitioned than U  in melt (i.e. in the crust)
than in solid (i.e. in the mantle). Consequently,  one expects that:
(Th/U)$_{Crust}>$ 3.8 (probably $4.5\div 5.5$), (Th/U)$_{Mantle}<$ 3.8 (probably $2\div 3$).
These expectations are confirmed by geochemical measurements in the crust and in the
upper mantle. Since
 geoneutrino experiments (including KamLAND) are dominated by the crust contribution,  they should 
then observe (Th/U)$_{Crust-dominated}>$ 3.8. Combining this datum with Crust and upper Mantle
 sampling, mass balance arguments \cite{Turcotte} can 
	be used to evaluate the  (Th/U)$_{Lower-Mantle}$  ratio. Therefore, we might infer mantle layering if 
(Th/U)$_{Lower-Mantle}$  is different from (Th/U)$_{Upper-Mantle}$.  

We analyze, now, how to attach errors to the  $^{232}$Th and $^{238}$U abundance estimates.
One possible approach \cite{Fiorentini} is to evaluate central values and errors from spread of 
published $^{232}$Th and $^{238}$U estimates (e.g., attach $\pm 3\sigma $ significance to 
extremal values). This  approach is good as a first guess, but is also affected by some limitations: 1) the published estimates
are often "duplicates" (i.e., they depend on each other); 2) without a criterion to discard obsolete estimates or unreliable outliers,
no progress is possible in reducing errors. As consequence, there is no  way  out than 
carefully sifting the available geo-literature, identifying virtues and problems of each estimate,
selecting the more reliable and complete ones, and evaluating from  scratch the uncertainties.
Error evaluation is not as common practice in geo-sciences as it is in particle physics, 
unfortunately.

Only recently (2003) a geochemical Earth model has appeared, in which input uncertainties are
well defined (although questionable in size) and propagated to output element abundances \cite{Palme}
through standard statistical techniques.
The corresponding Th and U estimates for the Bulk Silicate Earth (mantle plus crust) at $1\sigma $
can be expressed as:

\begin{center}
$[$\,Th\,$]$ = 83.5 (1$\pm $0.12) ppb\,,\\
$[$ U \,$]$ = 21.9 (1$\pm $0.12) ppb\,,\\
\end{center}
\noindent
with correlation $ \rho  =0.38$ (our provisional estimate). Consequently, we obtain the following
ratio:
\begin{center}
Th/U $\simeq $ 3.8 (1$\pm $0.14)  \hspace{0.5cm}   (for bulk Earth).
\end{center}
\noindent
This estimate can be refined through a more careful use of meteoritic (chondritic) data.
This is an important task since (Th/U)$_{Earth}$ plays a pivotal role in Mantle-Crust balances.
In fact,
the value of (Th/U)$_{Crust}>$ (Th/U)$_{Earth}$ (testable by KamLAND) might be used  to evaluate if
(Th/U)$_{Mantle}<$ (Th/U)$_{Earth}$. As emphasized before, this is a potential tool to test the difference between 
upper mantle and lower mantle.
This information must be folded with careful estimates of (Th/U) variations, both vertical 
(crust layers and mantle layers) and horizontal (crust types).

In principle, large amounts of data and constraints are available, but dedicated global studies
are still lacking. Interaction with geo-science community would be beneficial.

\section{Desiderata for KamLAND Data Analysis}

In this section, we analyze the information coming from the KamLAND experiment and 
comment on its 
role in future geo-neutrino data analysis.
A model-independent check that, e.g., the Th/U ratio in the crust is greater than 
the chondritic value,
requires that Th and U component  are left free in the analysis.
The current (binned) data can give only  very weak constraints, of course.
Since the statistics will be very low for quite some time, it is wise to use as much information as
possible. So, it would be useful to avoid the binning procedure, and tag each single event in energy (recoverable
from the current KamLAND plots).  

\begin{figure}[h]
\begin{center}
\epsfig{figure=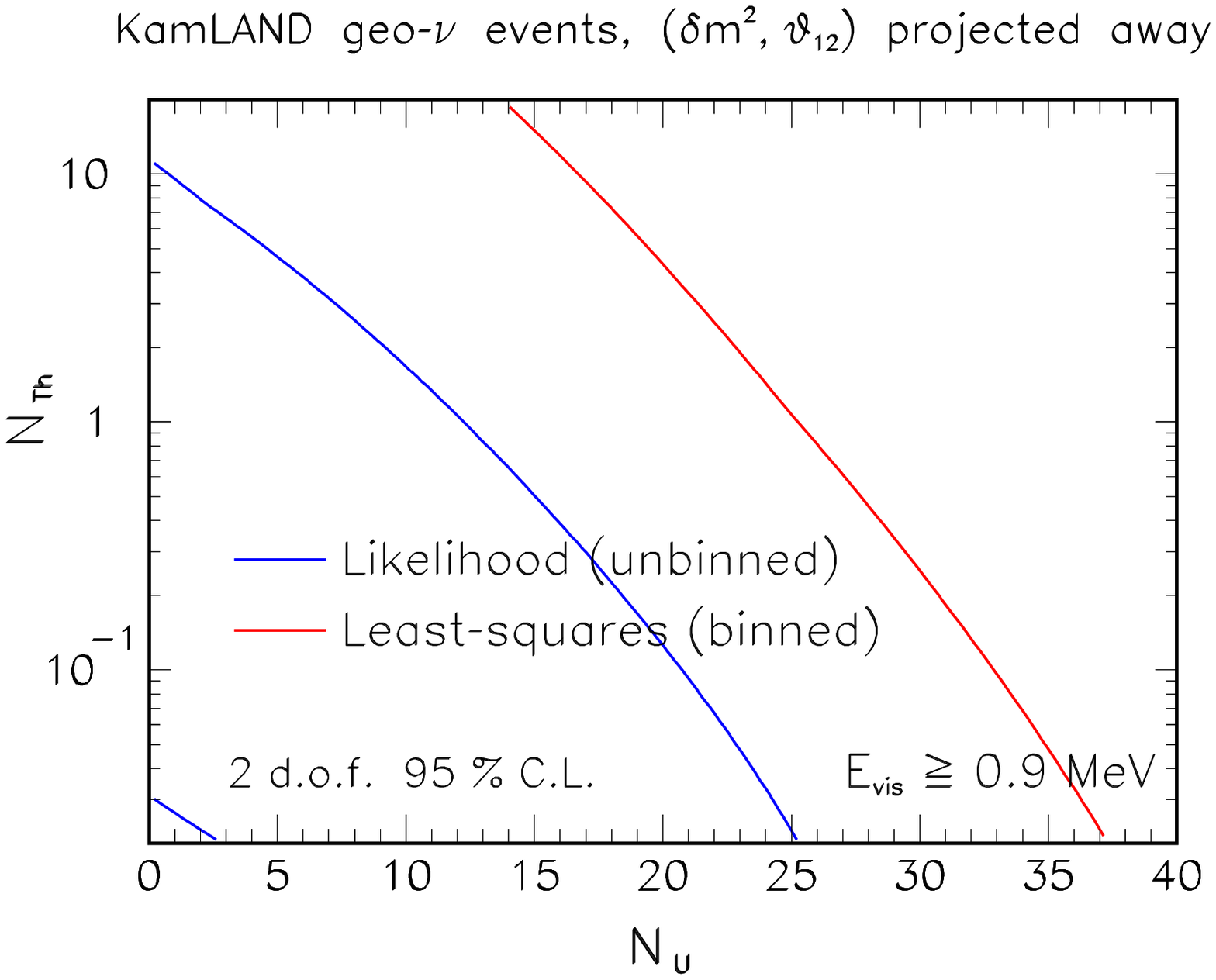,height=6.cm,width=8.5cm}

\vspace{0.7cm}

\epsfig{figure=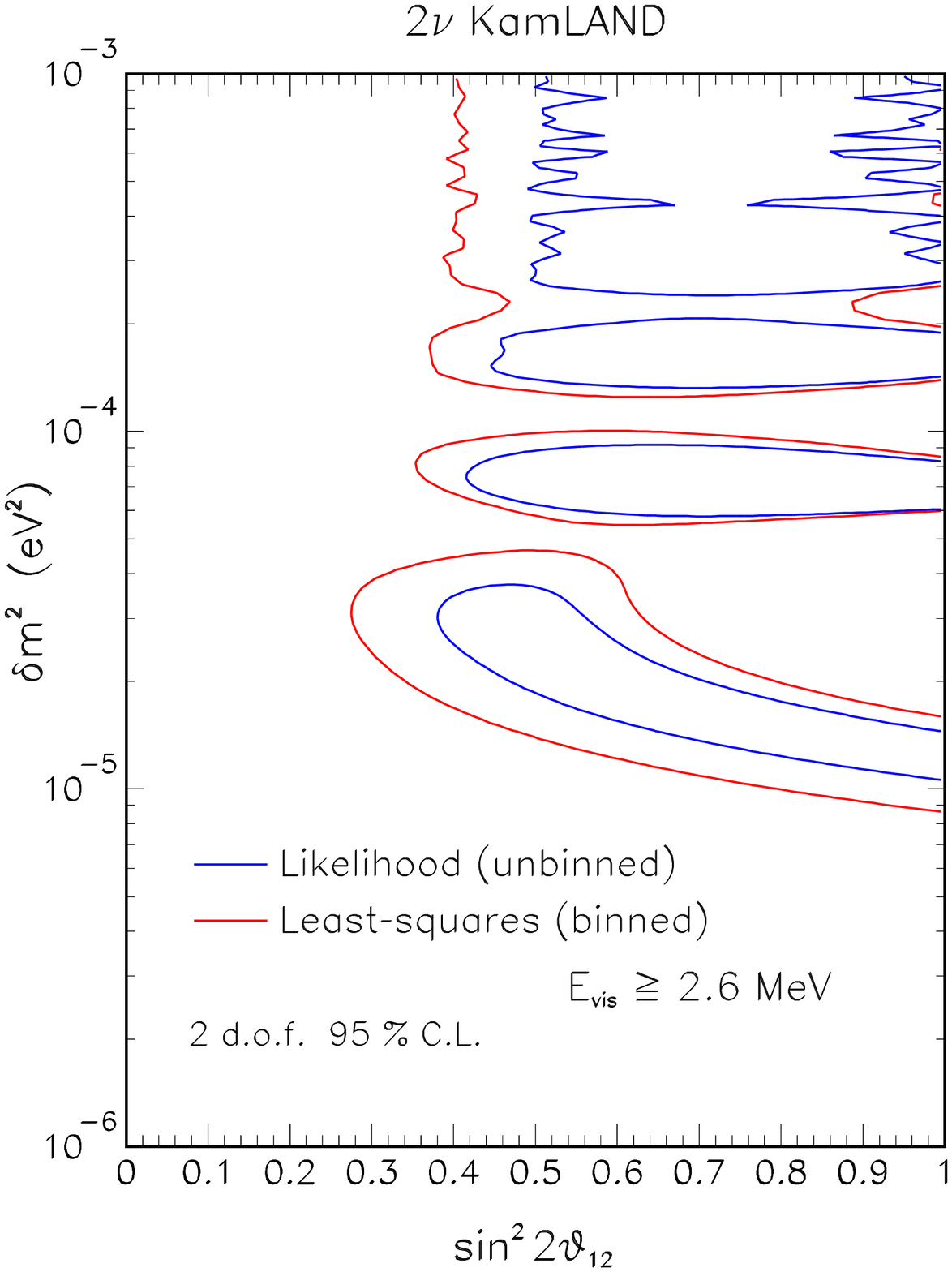,height=8.cm,width=6.5cm}		
\caption{Upper Panel: comparison of the 95\% C.L. regions obtained from the 
  Likelihood  analysis with the regions obtained from the Least-Squares analysis, projected onto the 
 plane of U and Th number of events. Lower-Panel: comparison of the 95\% C.L. regions obtained from the 
  Likelihood  analysis with the regions obtained from the Least-Squares analysis in the
 oscillation parameter space. \hspace{8.5 cm}
\label{1} }
\end{center}
\end{figure}

Fig.\,\ref{1} (upper panel) shows the iso-contours at 95\% C.L. in a four-parameters analysis,
projected onto the plane charted by the  U and Th number of $\bar{\nu}_{geo}$  events in KamLAND. 
The figure shows the comparison 
between Least-Squares (binned) and Likelihood (unbinned) analysis. It appears that unbinned analysis of KamLAND data (through maximum likelihood)
yelds thighter constraints on Th and U contributions. The loss of information implied by the 
binning of the data is non-negligible, and the likelihood analyis provides a more powerful method
to extract information from the KamLAND data. We conclude that it should be used as a default.
So, the first desideratum is that the KamLAND collaboration should provide the energies of each
event. We notice, as also shown in \cite{Schwetz}, that unbinned analyses provide better constraints not only on Th and U contributions
but even  on the oscillation parameters, as shown in Figure \ref{1} (lower panel). 

\begin{figure}[!th]
\begin{center}
\epsfig{figure=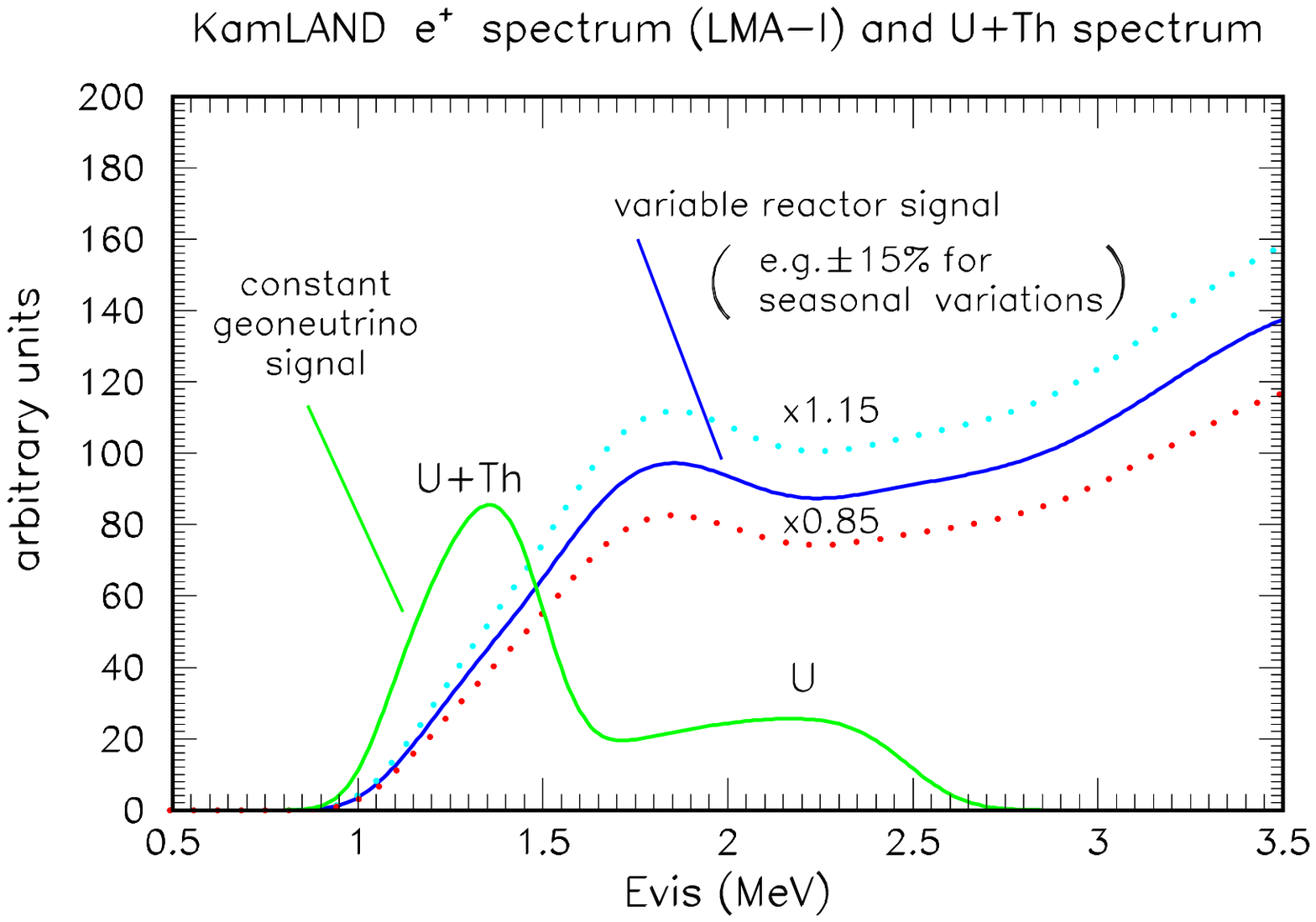,width=10.cm,height=7.cm}	
\caption{KamLAND positron spectrum (LMA-I) and geoneutrino spectrum (U and Th).
        The figure shows possible time variations (e.g. $\pm 15$\% for 
     seasonal variation) of the reactor component, and the time-independent trend of the 
		geo-neutrino component. \hspace{10.5 cm}
\label{2}}
 \end{center}
\end{figure}

A further and equally important requirement,
 is to use time information. The geo-neutrino flux is constant, while the flux
coming from reactors follows seasonal and/or occasional variations due to possible reactors' shut-downs
(known to
KamLAND Collaboration). Fig. \ref{2} shows the constant geoneutrino signal, superimposed to
the variable reactor signal. It can be seen that typical time variations of the reactor signal
can be as large as the contribution from $\bar{\nu}_{geo}$.  This is an additional handle to separate the constant geo-neutrino 
contribution from the variable reactor signal \cite{Lisi}, if an unbinned energy-time maximum likelihood
is used. As Figure \ref{2} shows, the main variation of the reactor signal corresponds to the 
U tail in the geoneutrino spectrum. Therefore, future data analyses with possibly lower experimental
threshold would be useful to evaluate the U component, and to discriminate U and Th contribution
in the total geoneutrino flux. The second desideratum is thus that, for each KamLAND event, energy plus time 
information should be released, and the reactors flux history should also be provided.

\newpage
\section{Future Prospects}
Concerning the future, the list of the desiderata must certainly include new experiments 
in geophysically different sites such as: Borexino, LENA, Sudbury,
Hawaii, Baksan. 


\begin{figure}[bh]
\begin{center}
\epsfig{figure=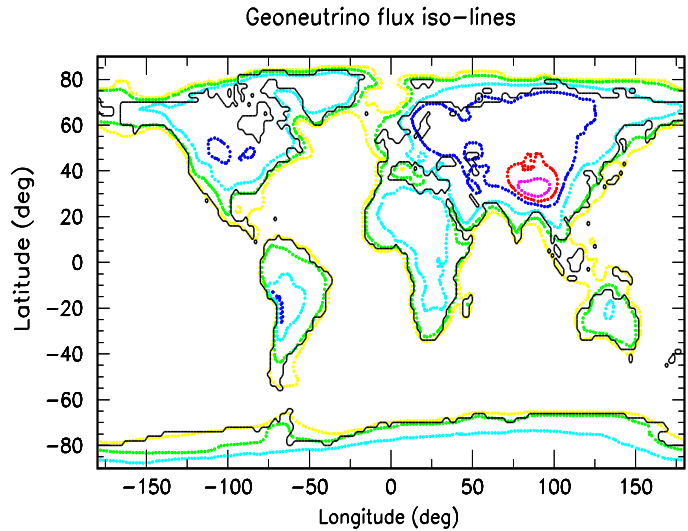,width=9.0cm,height=7.0cm}		
\caption{Geoneutrino flux iso-lines.
\label{3}}
 \end{center}
\end{figure}

A network of detectors located in different points of the Earth would be useful to obtain a complete and precise information about the different contribution to
the anti-neutrino fluxes. In fact,  detectors located in continental crust zones would give information
on the  dominant crust contribution, while detectors placed in  oceanic crust zones would help to measure
 the upper mantle contribution. Fig.\,\ref{3} shows  geoneutrino flux iso-lines with lowest fluxes in oceanic zones
and highest peaks at thick continental zones (e.g., the Himalaya chain).
 Several experiments are also needed to average out local uncertainties
in Th and U distributions, as explained in \cite{workinprogress}.

\section{Conclusions}

The KamLAND experiment will start a new field of geo-neutrino observations, and might be
followed by other similar experiments. 
Large uncertainties might hide the underlying (geo-) physics for quite some time, but steady progress
can be envisaged, if the particle physics and the geophysics communities identify common goals. 
In particular: 1) An effort can and should be made to characterize the geophysical and geochemical input in
``particle physics language'': central values, errors and correlations, e.g. on Th/U ratio.
 Missing input should be
identified and worked out by the two communities. 2) The few geo-neutrino events which will be collected by KamLAND (and possibly other experiments) 
in the future deserve our best analysis tools, to squeeze the maximum amount of information.
Time and energy tagging of each individual event can help to discriminate the 
geo-neutrino signal from the reactor component. Full publicity and full information of single events and 
single reactor history is essential to achieve this goal.

 \section*{Acknowledgments}

A.M. R. would like to thank the organizers of this conference for the kind hospitality
and the stimulating background that they provided, and for their interest in the topic of this talk.

\end{document}